\documentclass[preprint,aps,nofootinbib]{revtex4}
\usepackage{}
\usepackage{graphicx}%Include figure files
\usepackage{amsmath}
\usepackage{amsfonts}
\usepackage{amssymb}
\usepackage{color}%
\usepackage{dcolumn}% Align table columns on decimal point
\providecommand{\U}[1]{\protect\rule{.1in}{.1in}}

\definecolor{lightgray}{rgb}{.7,.7,.7}

\definecolor{red}{rgb}{1,0,0}

\definecolor{blue}{rgb}{0,0,1}

\newcommand{\f}{\begin{equation}}
\newcommand{\ff}{\end{equation}}
\newcommand{\fa}{\begin{eqnarray}}
\newcommand{\ffa}{\end{eqnarray}}

\begin{document}
\title{From Petrov-Einstein-Dilaton-Axion to Navier-Stokes equation in anisotropic model}

\author{Wen-Jian Pan $^{1}$}
\email{wjpan_zhgkxy@163.com}
\author{Yu Tian $^{2,4}$}
\email{ytian@ucas.ac.cn}
\author{Xiao-Ning Wu $^{3,4}$}
\email{wuxn@amss.ac.cn}

\affiliation{$^1$ Institute of theoretical physics,Beijing University of Technology,Beijing,100124,China\\
$^2$ School of Physics, University of Chinese Academy of Sciences, Beijing 100049, China\\
$^3$ Institute of Mathematics, Academy of Mathematics and System Science,
Chinese Academy of Sciences, Beijing 100190, China\\
$^4$  State Key Laboratory of Theoretical Physics, Institute of Theoretical Physics, Chinese Academy of Sciences, Beijing 100190, China}

\begin{abstract}
In this paper we generalize the previous works to the case that the near-horizon dynamics of the Einstein-Dilaton-Axion theory can be governed by the incompressible Navier-Stokes equation via imposing the Petrov-like boundary condition on hypersurfaces in the non-relativistic and near-horizon limit. The dynamical shear viscosity $\eta$ of such dual horizon fluid in our scenario, which isotropically saturates the Kovtun-Son-Starinet (KSS) bound, is independent of both the dilaton field and axion field in that limit.
\end{abstract}

\maketitle

\section {Introduction}
The correspondence between anti-de Sitter gravity and Conformal Field Theory (AdS/CFT)
proposed in \cite{Maldacena:1997re,Gubser:1998bc,Witten:1998qj,Aharony:1999ti}
provides a powerful tool on the connection between gravitational
physics in the bulk and hydrodynamics living on its boundary.
Since Damour firstly found that the gravitational excitations
of the black hole horizon behaved like a fluid\cite{Damour1979},
the hydrodynamical behavior of gravity has been extensively
studied in literature\cite{P-T,Jacobson:1995ab,PSS,KSS,B-L,I-L,
Bhattacharyya:2008kq,EFO,Padmanabhan10rp,Wilsonian,Heemskerk10hk,Faulkner10jy,
Bredberg:2011jq,Compere:2011dx,Cai,Bredberg:2011xw,Niu:2011gu,Matsuo11fk}.
In particular, recent progress on fluid/gravity duality in the context of AdS/CFT has shed more insightful light on relating the
Einstein's equation to the Navier-Stokes equation for a general
class of spacetime geometries\cite{Bhattacharyya:2008kq,EFO,Wilsonian}.
In this setup the gravitational fluctuations confined
in between the horizon and a finite cutoff at radius $r=r_c$,
can be mapped into a dual holographic fluid living on the cutoff
surface. Traditionally, directly disturbing the bulk metric
under the regularity condition of the horizon and fixing
the induced metric on the boundary, the correspondence
between gravitational dynamics in the bulk and hydrodynamics
on its boundary can be constructed successfully in the non-relativistic
long-wavelength expansion, and those dual hydrodynamical quantities
can be also explicitly read off via the standard procedure
in AdS/CFT dictionary, whose dependence on the
cutoff $r_c$ is viewed as the renormalization
group flow in the fluid\cite{Bredberg:2011jq,Compere:2011dx,Cai,
Bredberg:2011xw,Niu:2011gu,Matsuo11fk}.

Very remarkably, the gravity/fluid duality was firstly implemented by
imposing the Petrov-like condition on the cutoff
surface in the near horizon limit \cite{Lysov11xx}
instead of the regularity condition on the horizon in
Rindler spacetime. It has been shown that embedding a
hypersurface $\Sigma_c$ into a Rindler spacetime,
the gravitational fluctuation can be reduced exactly to
the incompressible Navier-Stokes equation living on
one lower dimensional flat spacetime.
More explicitly, in this approach keeping the induced metric fixed
and taking the extrinsic curvature as fundamental variables,
one directly required the extrinsic curvature
perturbations to satisfy the Petrov-like condition
such that in the non-relativistic limit and the near horizon limit
the continuous equation of the Brown-York tensor
can give rise to the
incompressible Navier-Strokes equation. In this sense,
the Petrov-like condition plays an important
holographic role on this correspondence. In contrast to
traditional approaches, this kind of setup is mathematically
much simpler and elegant, since it doesn't even need to
construct explicitly the metric perturbation
in the bulk, thus no need to solve the perturbed
Einstein equations in the bulk either. Due to this
 powerful condition, recently there have been greatly interesting extensions
in \cite{HL,Huang:2011kj,Zhang:2012uy,Wu:2013kqa,Wu:2013mda,
Ling:2013kua,Cai:2013uye,Cai:2014ywa,Cai:2014sua,Hao:2014xva,Hao:2015zxa}.

On the other hand,
an interesting conjecture, which said that the ratio of dynamical shear viscosity to entropy density was no less than ${1\over4\pi}$, was proposed
in \cite{Kovtun:2004de}. This is the so-called ``KSS bound".
However, this bound was later found to be violated in the anisotropic holographic plasma\cite{Rebhan:2011vd}. Since then, the problem of KSS bound violation has attracted a great
deal of attention\cite{Cheng:2014sxa,Cheng:2014qia,Ge:2014aza,
Jain:2014vka,Critelli:2014kra,Jain:2015txa} in the anisotropic gravitational
systems.

Motivated by the above progress, it should be interesting to ask how
about the gravity/fluid duality under the Petrov-like boundary condition in the context of an anisotropic gravitational
system. In this paper we will provide an answer to this question. It turns out that we can still obtain the standard Navier-Stokes equations under the non-relativistic and near horizon limit, but the anisotropy of the gravitational background only results in an anisotropy of the background pressure, while the ratio of dynamical viscosity to entropy density is still isotropic and saturates the KSS bound.

The rest of our paper is organized as follows. In Section 2 we briefly review
some important formulas. In Section 3 we will derive
the incompressible Navier-Strokes equations on a spatially
flat hypersurface from an anisotropic gravitational
system in detail. In Section 4 we will give a summary
and some discussions. In the appendix we present a detailed
calculation for the last term of the Petrov-like condition (\ref{petrov2}).

\section{some important formulas}
In this section, here we would like to review some important relations
that play the role of bridge on the gravity/fluid duality. Let us
start with the gravitational side.
Firstly, we naturally require p+2 dimensional spacetime geometry
to satisfy the standard Einstein theory:
\begin{eqnarray}\label{EE}
 G_{\mu\nu}=-{\Lambda}g_{\mu\nu}+T_{\mu\nu},\quad{\mu},{\nu}=0,{\ldots},p+1,
\end{eqnarray}
where $g_{\mu\nu}$ is a metric of the $p+2$ dimensional spacetime,
$\Lambda$ is a cosmological constant and $T_{\mu\nu}$ is energy momentum
tensor in the bulk.
Secondly, in order to discuss the behavior of dual fluid,
in the $p+2$ dimensional bulk space one needs to embed
a $p+1$ dimensional timelike hypersurface $\Sigma_c$
 with a induced metric $\gamma_{ab}$, whose extrinsic curvature
$K_{ab}$ should satisfy the $p+1$ ``momentum constraints"
 \begin{eqnarray}
 D^a(K_{ab}-{\gamma_{ab}}K)=T_{\mu b}n^{\mu},\label{mc}
 \end{eqnarray}
 as well as the ``Hamiltonian constraint"
 \begin{eqnarray}\label{HC1}
 {^{p+1}R}+{K_{ab}}{K^{ab}}-{K^2}-{2\Lambda} = -2T_{\mu\nu}n^{\mu}n^{\nu},
\end{eqnarray}
where $D_a$ is compatible with the induced metric on $\Sigma_c$,
namely $D_{a}{\gamma_{bc}}=0$, $K$ is the trace of extrinsic
curvature and $n^{\mu}$ is the unit normal to $\Sigma_c$.

For imposing Petrov-like condition on this cutoff surface
one must decompose the p+2 dimensional Weyl tensor into those
p+1 dimensional quantities in terms of the intrinsic curvature,
extrinsic curvature and induced metric on the hypersurface.
The framework has been specifically introduced
in previous literature\cite{Lysov11xx,HL,Huang:2011kj},
 \begin{eqnarray}\label{Weyl1}
C_{abcd}= ^{p+1}R_{abcd}+ K_{ad}K_{bc}- K_{ac}K_{bd}  + {2\Lambda-2T\over{p(p+1)}}(\gamma_{ad}\gamma_{bc}-\gamma_{ac}\gamma_{bd})\nonumber \\
    -{1\over p}{\gamma_a}^{\alpha}{\gamma_b}^{\beta}{\gamma_c}^{\gamma}{\gamma_d}^{\delta}
    (g_{\alpha\gamma}T_{\delta\beta}-g_{\alpha\delta}T_{\gamma\beta}-g_{\beta\gamma}T_{\delta\alpha}
    +g_{\beta\delta}T_{\gamma\alpha}),\nonumber \\
C_{abc(n)}= D_a K_{bc} - D_b K_{ac}  -{1\over p}{\gamma_a}^{\alpha}{\gamma_b}^{\beta}{\gamma_c}^{\gamma}n^{\delta}
    (g_{\alpha\gamma}T_{\delta\beta}-g_{\alpha\delta}T_{\gamma\beta}-g_{\beta\gamma}T_{\delta\alpha}
    +g_{\beta\delta}T_{\gamma\alpha}),\nonumber \\
C_{a(n)c(n)}=  K K_{ac} - {K_a}^b K_{bc} +{\gamma_a}^{\alpha}{\gamma_c}^{\gamma}R_{\alpha\gamma}
   - \ ^{p+1}R_{ac} -{2(\Lambda-T)\over{p(p+1)}}\gamma_{ac}\nonumber \\
    -{1\over p}{\gamma_a}^{\alpha}{\gamma_c}^{\gamma}n^{\beta}n^{\delta}(g_{\alpha\gamma}T_{\delta\beta}
    -g_{\alpha\delta}T_{\gamma\beta}-g_{\beta\gamma}T_{\delta\alpha}+g_{\beta\delta}T_{\gamma\alpha}).
\end{eqnarray}
 Thus the Petrov-like
boundary condition on $\Sigma_c$ is defined as
\begin{eqnarray} \label{petrov}
C_{(\ell)i(\ell)j}=\ell^{\mu}{m_i}^{\nu}\ell^{\alpha}{m_j}^{\beta}C_{\mu\nu\alpha\beta}=0,
\end{eqnarray}
where $p+2$ Newman-Penrose-like vector fields satisfy the following
relations
\begin{eqnarray}
\ell^2=k^2=0,\ \ (k,\ell)=1,\ \ (k,m_i)=(\ell,m_i)=0,\ \
(m^i,m_j)={\delta^i}_j,
\end{eqnarray}
where $\gamma_{ab}=g_{ab}-n_an_b$, $C_{abc(n)}=C_{abc\mu}n^{\mu}$.
In the absence of matter field, the traceless Petrov-like boundary
condition on $\Sigma_c$ actually causes $p(p+1)/2-1$ constraints
on the extrinsic curvature such that it can reduce exactly the $(p+1)(p+2)/2$ degrees of freedom of
extrinsic curvature to p+2 unconstrained variables which can be viewed
as the energy density, pressure and velocity fields of the dual fluid
living on the cutoff surface. The Hamiltonian constraint becomes a equation
of state linking the energy density to pressure of such dual fluid, while
the $p+1$ momentum constraints govern the evolution of the dynamics of gravity
which is regarded as a fluid living on hypersurface.

In the presence of matter field, on the surface we
generally need further to introduce some appropriate
boundary condition for matter field so that the total
degree of freedom can correctly present the dual
hydrodynamical behavior. For vacuum case of Einstein theory,
it can be governed by the initial-boundary value problem (IBVP)
\cite{Friedrich:1998xt}. Based on the idea by Friedrich and Nagy, we
can see that Petrov-like boundary condition can be viewed
as the free boundary data of IBVP of vacuum Einstein system.
This hint gives us a guideline for searching a suitable
boundary condition for matter field.

\section{Navier-Stokes Equations in the anisotropic spacetime}
In this section, employing the Petrov-like boundary condition
on the cutoff surface embedded in a five-dimensional anisotropic
spacetime, we will explicitly demonstrate how to derive
the incompressible Navier-Stokes Equations from the anisotropic linear
axion model under the near horizon and non-relativistic limit.
As showed in \cite{Cheng:2014qia,Mateos:2011ix,Mateos:2011tv},
the action of the Einstein-Dilaton-Axion theory can be written as
\begin{equation}\label{action}
S =\int d^5x\sqrt{-g} [\frac{1}{2\kappa^2}(R+12)-{1\over2}(\partial\phi)^2-{1\over2}e^{2\phi}(\partial\chi)^2],
\end{equation}
where $2\kappa^2=16\pi G_5$ is the five-dimensional gravitational coupling,
$\phi$ and $\chi$ are the dilaton field and the axion field, respectively.
Here we have set the cosmological constant scale $L=1$.
Then the equations of motion for the axion, the dilaton and the gravitational field
can be presented respectively
\begin{eqnarray}
 \nabla_{\mu}(e^{2\phi}\nabla^{\mu}\chi)=0,\\
\nabla_{\mu}\nabla^{\mu}\phi-e^{2\phi}(\partial\chi)^2=0,\\
R_{\mu\nu}-\partial_{\mu}\phi\partial_{\nu}\phi-e^{2\phi}\partial_{\mu}\chi\partial_{\nu}\chi+4g_{\mu\nu}=0.
\end{eqnarray}
Here we have used $\kappa^2=8\pi G_5=1$. We consider such spacetime geometry\cite{Cheng:2014qia,Mateos:2011ix,Mateos:2011tv} preserving
rotational invariance in the $x$-$y$ plane, which can be written as
\begin{equation}
    {ds^2_5} = e^{-\frac{\phi}{2}}[-f(r)B(r)dt^2+2\sqrt{B(r)}drdt+r^2dx^2+r^2dy^2+r^2H(r)dz^2],
\end{equation}
where $\phi,B,H$ and $f(r)\equiv r^2 F(r)$ only depend on the radial
coordinate $r$. $\chi$ is a linear function, namely $\chi=az$, where $a$ is a constant. At the horizon of this geometry $f(r_h)$ should
be vanishing, namely $f(r_h)=0$, which is equivalent to $F(r_h)=0$.
 $H(r)$ is related to the dilaton field $\phi$, namely $H(r)=e^{-\phi(r)}$.
When $H(r)=1$, this spacetime has spatial isotropy.
Otherwise, it is anisotropic. The relevant anisotropic solutions
have appeared in \cite{Cheng:2014qia,Mateos:2011ix,Mateos:2011tv}.
Thus here we do not try to repeat it in detail, whereas
mainly focus on the hydrodynamical behavior of the gravity.
The hypersurface $\Sigma_{c}$ located by $r=r_c$ outside
the horizon of this geometry can be naturally introduced by
\begin{eqnarray}
    ds^2_4    &=&     -f(r_c)B(r_c)e^{-\frac{1}{2}\phi(r_c)}dt^2 + r^2_ce^{-\frac{1}{2}\phi(r_c)}dx^2+r^2_ce^{-\frac{1}{2}\phi(r_c)}dy^2
    +r^2_cH(r_c)e^{-\frac{1}{2}\phi(r_c)}dz^2 \nonumber \\
    &=&     -{dx^0}^2 +r^2_ce^{-\frac{1}{2}\phi(r_c)}dx^2+r^2_ce^{-\frac{1}{2}\phi(r_c)}dy^2
    +r^2_cH(r_c)e^{-\frac{1}{2}\phi(r_c)}dz^2,
\end{eqnarray}
where $\sqrt{f(r_c)B(r_c)e^{-\frac{1}{2}\phi(r_c)}}t = x^0$.
Obviously the hypersurface embedded is intrinsic flat.
To exhibit explicitly the non-relativistical behavior
of dual hydrodynamics on this hypersurface, we further need to introduce
a parameter $\lambda$ by rescaling the time coordinate
$\lambda x^0=\tau$. Thus the above induced metric can be rewritten as
\begin{equation}
{ds^2}_{p+1}=-\frac{1}{\lambda^2}d\tau^2 + r^2_ce^{-\frac{1}{2}\phi(r_c)}dx^2+r^2_ce^{-\frac{1}{2}\phi(r_c)}dy^2
    +r^2_cH(r_c)e^{-\frac{1}{2}\phi(r_c)}dz^2,
\end{equation}
Later, one can see that both the non-relativistical limit
and the near horizon limit will be implemented by identifying
the parameter $\lambda$ with the location of
the hypersurface, namely $r_c-r_h=(\alpha\lambda)^2$
such that taking $\lambda\rightarrow 0$ means
these limits can be achieved simultaneously. Taking $\lambda\to 0$ limit implies that
the hypersurface is highly accelerated, which is thought of as the large mean curvature. It is easily checked that the background quantities of
the components of extrinsic curvature defined
on the hypersurface $\Sigma_{c}$ have following
forms in the coordinate$(\tau,x^i)$
\begin{eqnarray}
    {K^{\tau(B)}}_{\tau}&=&e^{\frac{1}{4}\phi_c}[\frac{f^{\prime}_c}{2\sqrt{f_c}}
    +\frac{\sqrt{f_c}B^{\prime}_c}{2B_c}-\frac{\sqrt{f_c}\phi^{\prime}_c}{4}] \nonumber\\
    {K^{\tau(B)}}_i&=&0 \nonumber\\
    {K^{2(B)}}_2&=&{K^{1(B)}}_1=e^{\frac{1}{4}\phi_c}(\frac{\sqrt{f_c}}{r_c}-\frac{\sqrt{f_c}\phi^{\prime}_c}{4}) \nonumber\\
    {K^{3(B)}}_3&=&e^{\frac{1}{4}\phi_c}(\frac{\sqrt{f_c}}{r_c}-\frac{\sqrt{f_c}\phi^{\prime}_c}{4}+\frac{\sqrt{f_c}H^{\prime}_c}{2H_c}) \nonumber\\
    K&=&e^{\frac{1}{4}\phi_c}[\frac{f^{\prime}_c}{2\sqrt{f_c}}
    +\frac{\sqrt{f_c}B^{\prime}_c}{2B_c}+\frac{3\sqrt{f_c}}{r_c}-{\sqrt{f_c}\phi^{\prime}_c}+\frac{\sqrt{f_c}H^{\prime}_c}{2H_c}],
\end{eqnarray}
 where the prime $\prime$ denotes derivative with respect to $r$. 
Here, for the convenience, we have abbreviated the background terms,
namely $\Phi(r_c)\equiv\Phi_c$, where $\Phi(r_c)$ includes $f(r_c)$, $\phi(r_c)$, $B(r_c)$ and $H(r_c)$ above.
Furthermore, in order to describe definitely the perturbation effect of gravity,
as appearing in \cite{HL}, we still take the Brown-York stress tensor
as fundamental variable, which is defined as
\begin{equation}
{t^a}_b={\delta^a}_b K-{K^a}_b,
\end{equation}
In coordinate$(\tau,x^i)$, we can further rewrite the components
of extrinsic curvature and its trace in terms of their corresponding
Brown-York tensors ,
\begin{equation} \label{EC}
{{K^\tau}_\tau}={t\over{3}}-{t^\tau}_\tau,\ \ \ \ \
{{K^\tau}_i}=-{t^\tau}_i,\ \ \ \ \
{{K^i}_j}=-{t^i}_j+{\delta^i}_j{t\over{3}},\ \ \ \ \
K=\frac{t}{3}.
\end{equation}
Now we can start to investigate the hydrodynamical behavior of
the gravity on the cutoff surface
in the near horizon limit and non-relativistic limit. Note that in contrast
to the conventional perturbation method that uses the metric expansion
to solve the perturbation Einstein equations, and then governs the
Brown-York tensor dynamical behavior identified with the energy
momentum tensor of hydrodynamics on hypersurface, here we take the Brown-York tensors
as the fundamental variables and consider directly its fluctuations
on the cutoff surface, without solving the perturbation gravitational equations,
while keeping the intrinsic induced metric of the
surface fixed. Thus we can expand their components in powers of $\lambda$ as
\begin{eqnarray} \label{ttau1}
    {t^\tau}_i       &=&    0 + \lambda{{t^\tau}_i}^{(1)} + \ldots  \nonumber\\
    {t^\tau}_\tau    &=&    e^{\frac{1}{4}\phi_c}[{3\over{r_c}}\sqrt{f_c}-\frac{3\sqrt{f_c}\phi^{\prime}_c}{4}+\frac{\sqrt{f_c}H^{\prime}_c}{2H_c}] + \lambda{{t^\tau}_\tau}^{(1)} + \ldots \nonumber\\
    {t^i}_j          &=&    e^{\frac{1}{4}\phi_c}[\frac{f^{\prime}_c}{2\sqrt{f_c}}
        +\frac{\sqrt{f_c}B^{\prime}_c}{2B_c}+\frac{2\sqrt{f_c}}{r_c}-\frac{3\sqrt{f_c}\phi^{\prime}_c}{4}
        +\frac{\sqrt{f_c}H^{\prime}_c}{2H_c}-{\delta^i}_3{\delta^3}_j\frac{\sqrt{f_c}H^{\prime}_c}{2H_c}]{\delta^i}_j
        +  \lambda {{t^i}_j}^{(1)} + \ldots \nonumber\\
    t\,\,            &=&    3e^{\frac{1}{4}\phi_c}[\frac{f^{\prime}_c}{2\sqrt{f_c}}
        +\frac{\sqrt{f_c}B^{\prime}_c}{2B_c}+\frac{3\sqrt{f_c}}{r_c}-{\sqrt{f_c}\phi^{\prime}_c}
        +\frac{\sqrt{f_c}H^{\prime}_c}{2H_c}] + \lambda t^{(1)}+\ldots,
\end{eqnarray}
where the Latin alphabets $i,j$ of the term
${\delta^i}_3{\delta^3}_j\frac{\sqrt{f_c}H^{\prime}_c}{2H_c}$
inside the square brackets in the third line of
the above equations runs as the Latin alphabets $i,j$
outside the square brackets, but the former does not join in
other behaviors such as contraction.
To obtain the perturbation behavior of gravity in the
near horizon limit, we need to expand the background
terms of the above stress tensors around the horizon in powers of $r_c-r_h$
identified with $\alpha^2\lambda^2$.
\begin{eqnarray}\label{BE}
f_c&=&f^{\prime}_h\alpha^2\lambda^2+\frac{f^{\prime\prime}_h}{2}\alpha^4\lambda^4+\ldots\nonumber\\
H_c&=&H_h+H^{\prime}_h\alpha^2\lambda^2+\ldots\nonumber\\
B_c&=&B_h+B^{\prime}_h\alpha^2\lambda^2+\ldots\nonumber\\
\phi_c&=&\phi_h+\phi^{\prime}_h\alpha^2\lambda^2+\ldots
\end{eqnarray}
one will see that this strategy unifying the non-relativistic
limit and near horizon limit plays an essential role in
deducing successfully the standard Navier-Strokes Equation.
Now we consider the specific form of ``Hamiltonian constraint"
in Eq.(\ref{HC1}), whose form is rewritten in terms of
the components of Brown-York stress tensor
\begin{equation}\label{HC2}
 {t^{\tau}}_{\tau}{t^{\tau}}_{\tau}+{t^n}_m{t^m}_n-\frac{t^2}{3}-
\frac{2}{\lambda^2}\gamma^{mn}{t^{\tau}}_{m}{t^{\tau}}_{n}=-12-2T_{\mu\nu}n^{\mu}n^{\nu}
\end{equation}
Note that all the indices of the physical quantities on the hypersurface
here are lowered or raised with $\gamma_{ab}$
and $\gamma^{ab}$. Plugging Eq.(\ref{BE}) and Eq.(\ref{ttau1}) into
Eq.(\ref{HC2}), we find that the leading term of the constraint
at the order $\lambda^{-2}$ vanishes automatically and the sub-leading one
gives rise to
\begin{equation}\label{HC3}
{t^{\tau}}_{\tau}^{(1)}=-2e^{-\frac{1}{4}\phi_h}\gamma^{(0)mn}{t^{\tau}}_m^{(1)}{t^{\tau}}_n^{(1)}-
3e^{\frac{1}{4}\phi_h}(\frac{f^{\prime}_h}{r_h}-
\frac{3f^{\prime}_h}{4}\phi^{\prime}_h)+e^{-\frac{1}{4}\phi_h}(12-\frac{a^2e^{\frac{7\phi_h}{2}}}{r^2_h})
\end{equation}
where $\gamma^{(0)mn}\equiv\gamma^{mn}(r_h)$ and $H(r)=e^{-\phi(r)}$ have been used.
Now we turn to considering Petrov-like boundary condition on
hypersurface. After choosing $3+2$ Newman-Penrose-like vector fields,
\begin{eqnarray}\label{NPL}
\sqrt{2}\ell=\partial_0-n,\ \ \sqrt{2}k=-\partial_0-n,\ \ m_1=\frac{e^{\frac{1}{4}\phi}}{r}\partial_1,
\ \ \ m_2=\frac{e^{\frac{1}{4}\phi}}{r}\partial_2,\ \ m_3=\frac{e^{\frac{1}{4}\phi}}{r\sqrt{H}}\partial_3,
\end{eqnarray}
the Petrov-like boundary condition (\ref{petrov}) can
be generally presented as
\begin{equation}\label{petrov1}
C_{0i0j}+C_{0ij(n)}+C_{0ji(n)}+C_{i(n)j(n)}=0
\end{equation}
To obtain conveniently the dynamical behavior of this geometry,
making use of Eq.(\ref{Weyl1}) we can rewrite explicitly
the condition in terms of Brown-York stress tensor as
\begin{eqnarray}\label{petrov2}
{t^\tau}_{\tau}{t^k}_j+{2\over\lambda^2}\gamma^{ki}{{t^\tau}_i}{{t^\tau}_j}
-{t^k}_i{t^i}_j+2\lambda\partial_{\tau}({t\over 3}{\delta^k}_j-{t^k}_j)+({t\over 3})^2{\delta^k}_j\nonumber\\
-{1\over\lambda}\gamma^{ki}(\partial_j{t^{\tau}}_i+\partial_i{t^{\tau}}_j)
-{t^{\tau}}_{\tau}({t\over 3}){\delta^k}_j+{B^k}_j=0,
\end{eqnarray}
where
\begin{eqnarray}\label{RT}
{B^k}_j\equiv\gamma^{ki}{\gamma_i}^{\alpha}{\gamma_j}^{\beta}R_{\alpha\beta}-\frac{1}{3}(\lambda^2{\delta^k}_jT_{\tau\tau}
+{\delta^k}_jT_{\alpha\beta}n^{\alpha}n^{\beta}-2\lambda{\delta^k}_jT_{\tau\alpha}n^{\alpha}).
\end{eqnarray}
The details of the calculation can be found in the Appendix. Substituting Eq.(\ref{ttau1}) and Eq.(\ref{BE}) into the above
equation (\ref{petrov2}), we find that the background as a leading term
satisfies automatically the Petrov-like condition
at the order of $1\over\lambda^2$:
\begin{equation}
\frac{f^{\prime}_h}{4\alpha^2\lambda^2}e^{\frac{1}{2}\phi_h}{\delta^k}_j-\frac{f^{\prime}_h}{4\alpha^2\lambda^2}e^{\frac{1}{2}\phi_h}{\delta^k}_j=0,
\end{equation}
and the perturbations of the gravity as sub-leading terms
at the order of $\lambda^0$, which gives rise to
\begin{eqnarray}\label{t1}
{t^k}_j^{(1)}=2e^{-\frac{1}{4}\phi_h}\gamma^{(0)ki}{t^{\tau}}_i^{(1)}{t^{\tau}}_j^{(1)}
+e^{\frac{1}{4}\phi_h}(\frac{f^{\prime}_h}{r_h}-\frac{f^{\prime}_h}{4}\phi^{\prime}_h
-{\delta^k}_3{\delta^3}_j\frac{f^{\prime}_h}{2}\phi^{\prime}_h){\delta^k}_j+{t^{(1)}\over 3}{\delta^k}_j\nonumber\\
-e^{-\frac{1}{4}\phi_h}\gamma^{(0)ki}(\partial_j{t^{\tau}}_i^{(1)}
+\partial_i{t^{\tau}}_j^{(1)})+e^{-\frac{1}{4}\phi_h}(-4{\delta^k}_j+{a^2e^{\frac{7}{2}\phi_h}\over r^2_h}\delta^{k3}{\delta^3}_j).
\end{eqnarray}
Here we have used $\frac{\sqrt{f^{\prime}_h}}{\alpha}=1$. In intrinsic flat hypersurface, since $T_{\mu b}n^{\mu}$ vanishes for choosing $b=\tau,i$,
the momentum constraint(\ref{mc}) in terms of Brown-York tensor ${t^a}_b$ reduces to be
\begin{equation}
\partial_a{t^a}_b=0.
\end{equation}
When $b=\tau$, we can directly derive the incompressible
condition from the above equation, which is
\begin{eqnarray}\label{Inp1}
O(\lambda^{-1}):& & \partial_k{\upsilon}^{k}=0.
\end{eqnarray}
When $b=j$,  utilizing Eq.(\ref{t1}), we can straightforwardly
derive the standard Navier-Stokes equations,
\begin{eqnarray}
\partial_\tau {\upsilon_j} + \upsilon^k\partial_k \upsilon_j -
 \nu\partial^2 \upsilon_j + \partial_j P_{\perp} = 0,\quad (j=1,2)\label{sdns1}\\
\partial_\tau {\upsilon_3} + \upsilon^k\partial_k \upsilon_3 -
 \nu\partial^2 \upsilon_3 + \partial_3 P_{\parallel} = 0.\quad (j=3)\label{sdns2}
\end{eqnarray}
Here we have defined the transverse and longitudinal pressures as
\begin{eqnarray}
P_{\perp} &=& 2e^{-\frac{1}{4}\phi_h}[\frac{t^{(1)}}{3}+e^{\frac{1}{4}\phi_h}(\frac{f^{\prime}_h}{r_h}-\frac{f^{\prime}_h}{4}\phi^{\prime}_h)-4e^{-\frac{1}{4}\phi_h}],\label{pperp}\\
P_{\parallel} &=& 2e^{-\frac{1}{4}\phi_h}[\frac{t^{(1)}}{3}+e^{\frac{1}{4}\phi_h}(\frac{f^{\prime}_h}{r_h}
-\frac{3f^{\prime}_h}{4}\phi^{\prime}_h)+e^{-\frac{1}{4}\phi_h}(\frac{a^2e^{\frac{7}{2}\phi_h}}{r^2_h}-4)]\label{ppara},
\end{eqnarray}
respectively.
 It is worth noting that, in the context of AdS/CFT,
the background pressures in the dual fluid contain some important background information in the bulk.
The different pressures in the dual Navier-Stokes equations (\ref{sdns1}) and (\ref{sdns2}),
in some sense, reversely indicate that the dual spacetime is anisotropic,
which distinguishes from the situation with identical pressures that mean the dual spacetime is isotropic.
In this sense, the background pressures are non-trivial. However,
the difference between the transverse pressure $P_{\perp}$ and the longitudinal one $P_{\parallel}$,
as shown in Eqs.(\ref{pperp}) and (\ref{ppara}), is just a constant. Moreover, from Eqs.(\ref{pperp}) and (\ref{ppara}), it is easy to
find that the background pressures contribute nothing to the dynamical behavior in the dual hydrodynamics, since their spatial derivatives vanish automatically.
As a consequence, we can drop the background pressures terms so that the usual Navier-Stokes equation can be still presented as
\begin{equation}
\partial_\tau {\upsilon_j} + \upsilon^k\partial_k \upsilon_j -
 \nu\partial^2 \upsilon_j + \partial_j P = 0,\quad (j=1,2,3)
\end{equation}
where the pressure $P$ has been identified with $ 2e^{-\frac{1}{4}\phi_h}\frac{t^{(1)}}{3}$.
In addition, we have also identified ${t^{\tau}}_j^{(1)}=\frac{1}{2}e^{\frac{1}{4}\phi_h}\upsilon_j$
and the kinematic shear viscosity $\nu=e^{-\frac{1}{4}\phi_h}$ above.
In particular, the ratio of dynamical viscosity to entropy density is
\begin{equation}
\frac{\eta}{s}=\frac{\nu\rho}{s}={\frac{1}{2}e^{\frac{1}{4}\phi_h} e^{-\frac{1}{4}\phi_h}\over\frac{1}{4G}}=2G=\frac{1}{4\pi}.
\end{equation}
Here we have used $8\pi G=1$, and the entropy density $s=\frac{1}{4G}$.
The above equation indicates that under the non-relativistic and near-horizon limit the dynamical viscosity of this dual fluid is still isotropic and saturates the KSS bound\cite{Kovtun:2004de}, even in the anisotropic holographic setup considered here.

\section{Summary and Discussions}
In this paper we have generalized the previous works\cite{Lysov11xx,HL,Huang:2011kj}
to the case in which the dynamical behavior of the Einstein-Dilaton-Axion theory can be governed by the incompressible
Navier-Stokes equations via imposing the Petrov-like boundary
condition on hypersurface in the non-relativistic limit as well as
in the near horizon limit, such that the holographic nature
and the elegance of the Petrov-like condition have
been further disclosed. Here requiring that the Petrov-like
condition holds on the cutoff surface, while keeping
the induce metric on this surface fixed, we have demonstrated
that in contrast with the Navier-Stokes equation with
unit kinematic shear viscosity in the previous works, the kinematic shear
viscosity of such fluid equation in our scenario is
related to the value of the dilaton field on the horizon. However, the ratio of dynamical shear viscosity to entropy density is still the constant $\frac{1}{4\pi}$,
although the anisotropic effect has been considered.
This likely means that such boundary condition under
the large mean curvature limit sustains the KSS bound.
In addition, the anisotropic background spacetime gives rise to
the anisotropy of the background pressures, which distinguishes from the isotropic case
that leads to the same background pressure. The difference of dual hydrodynamic pressures
between the transversal and the longitudinal
to the anisotropic direction is only constant which is given by the gradient
of the axion field and the relevant values of the dilaton field on the horizon.
 Since the spatial derivatives of the background pressure terms vanish automatically,
they do not affect the dynamical effect of such dual fluid. As a result,
we can ignore these constant pressure terms and redefine the pressure in dual hydrodynamics such that
the usual Navier-Stokes equation can be still obtained.

In standard approaches in AdS/CFT, the (dynamical) shear viscosity of the boundary dual fluid in anisotropic setups generally becomes a symmetric tensor with different eigenvalues in anisotropic directions, which violates the KSS bound in certain directions\cite{Rebhan:2011vd,Jain:2014vka}. Our result for the horizon fluid is more like that considered in \cite{DG1,DG2}, where the ratio of dynamical shear viscosity to entropy density is always $\frac{1}{4\pi}$ in spite of the anisotropy of the holographic setup. Nevertheless, a deep understanding of the relationship between these formalisms is still lacking.

On the other hand, it should be an interesting problem that
adopting the traditional non-relativistic long-wavelength
limit, how about the hydrodynamic behaviors of gravity at finite cutoff surfaces
and the corresponding transport coefficients in such anisotropic systems. The systems at cutoff surfaces can interpolate between the horizon fluid and the boundary CFT, which is then related to the understanding of the different results of shear viscosities mentioned above.
This aspect is left for future works.

\begin{acknowledgments}
 We would like to thank Yi Ling for the useful discussion and suggestions. We are also grateful to the anonymous referee for helpful suggestions.
 This work is partly supported by the National Natural Science Foundation of China (Grant Nos. 11175245, 11575286 and 11475179).
\end{acknowledgments}

\section*{Appendix}
In this Appendix we need to present that the Ricci tensor
and the momentum energy tensor in bulk are how to contribute
the Petrov-like condition in detail. From the variation
of the action (\ref{action}), the momentum energy tensor
can be given by
\begin{equation}
 T_{\mu\nu}=\partial_{\mu}\phi\partial_{\nu}\phi-{1\over2}g_{\mu\nu}g^{\alpha\beta}\partial_{\alpha}\phi\partial_{\beta}\phi
+e^{2\phi}\partial_{\mu}\chi\partial_{\nu}\chi-{1\over2}e^{2\phi}g_{\mu\nu}g^{\alpha\beta}\partial_{\alpha}\chi\partial_{\beta}\chi,
\end{equation}
where $\chi$ is a linear Axion field, namely $\chi=az$ and $a$ is constant.
Using the above equation, the all components of this tensor
can be straightforwardly calculated as
\begin{eqnarray}
T_{tt}&=&-[{1\over2}g_{tt}g^{rr}(\partial_r\phi)^2+{1\over2}e^{{5\over2}\phi}g_{tt}\frac{a^2}{r^2H(r)}],\\
T_{tr}&=&-[{1\over2}g_{tr}g^{rr}(\partial_r\phi)^2+{1\over2}e^{{5\over2}\phi}g_{tr}\frac{a^2}{r^2H(r)}],\\
T_{ti}&=& 0,\\
T_{rr}&=& (\partial_r\phi)^2,\\
T_{ri}&=& 0,\\
T_{ij}&=&-[{1\over2}g_{ij}g^{rr}(\partial_r\phi)^2+{1\over2}e^{{5\over2}\phi}g_{ij}\frac{a^2}{r^2H(r)}-e^{2\phi}a^2{\delta^3}_i{\delta^3}_j],\\
T&\equiv& g^{\mu\nu}T_{\mu\nu}= -[{3\over2}g^{rr}(\partial_r\phi)^2+{3\over2}e^{{5\over2}\phi}\frac{a^2}{r^2H(r)}].
\end{eqnarray}
The Einstein equation (\ref{EE}) in the dilaton-axion model can be rewritten as
\begin{equation}
 R_{\mu\nu}=\frac{-T-12}{3}g_{\mu\nu}+T_{\mu\nu}.
\end{equation}
Imposing the above equations, Eq.(\ref{RT}) can be presented as
\begin{equation}
{B^k}_j=-4{\delta^k}_j+e^{{5\over2}\phi_c}\frac{a^2}{r_c^2H_c}\delta^{k3}{\delta^3}_j
-\frac{1}{3}{\delta^k}_jf_ce^{\frac{\phi_c}{2}}(\partial_r\phi_c)^2.
\end{equation}
Here we have used the result
\begin{equation}
T_{\tau\mu}n^{\mu}=0.
\end{equation}


\begin{thebibliography}{99}
\bibitem{Maldacena:1997re}
  J.~M.~Maldacena,
  %``The Large N limit of superconformal field theories and supergravity,''
  Int.\ J.\ Theor.\ Phys.\  {\bf 38}, 1113 (1999)
  [Adv.\ Theor.\ Math.\ Phys.\  {\bf 2}, 231 (1998)]
  [hep-th/9711200].
  %%CITATION = HEP-TH/9711200;%%

\bibitem{Gubser:1998bc}
  S.~S.~Gubser, I.~R.~Klebanov and A.~M.~Polyakov,
  %``Gauge theory correlators from noncritical string theory,''
  Phys.\ Lett.\ B {\bf 428}, 105 (1998)
  [hep-th/9802109].
  %%CITATION = HEP-TH/9802109;%%

\bibitem{Witten:1998qj}
  E.~Witten,
  %``Anti-de Sitter space and holography,''
  Adv.\ Theor.\ Math.\ Phys.\  {\bf 2}, 253 (1998)
  [hep-th/9802150].
  %%CITATION = HEP-TH/9802150;%%

\bibitem{Aharony:1999ti}
  O.~Aharony, S.~S.~Gubser, J.~M.~Maldacena, H.~Ooguri and Y.~Oz,
  %``Large N field theories, string theory and gravity,''
  Phys.\ Rept.\  {\bf 323}, 183 (2000)
  [hep-th/9905111].
  %%CITATION = HEP-TH/9905111;%%
  %3413 citations counted in INSPIRE as of 31 Mar 2015

\bibitem{Damour1979}
T. Damour, (1979), Quelques propri¡äet¡äes m¡äecaniques,
¡äelectromagn¡äetiques, thermodynamiques et quantiques des trous
noirs, Th`ese de doctorat d¡¯¡äEtat, Uni- versit¡äe Paris 6.
(available at http://www.ihes.fr/$\sim$ damour/Articles/). T.
Damour, (1982), Surface effects in black hole physics, in
Proceedings of the Second Marcel Grossmann Meeting on General
Relativity, Ed. R. Ruffini, North Holland , p. 587.

\bibitem{P-T}R.H. Price and K.S. Thorne, %\textquotedblleft{}Membrane
%viewpoint on black holes: properties and evolution of the stretched
%horizon,\textquotedblright{}
Phys. Rev. D 33, 915 (1986).

%\cite{Jacobson:1995ab}
\bibitem{Jacobson:1995ab}
  T.~Jacobson,
  %``Thermodynamics of space-time: The Einstein equation of state,''
  Phys.\ Rev.\ Lett.\   75, 1260 (1995) [arXiv:gr-qc/9504004].
  %%CITATION = PRLTA,75,1260;%%

\bibitem{PSS}G. Policastro, D.T. Son and A.O. Starinets, %\textquotedblleft{}The
%shear viscosity of strongly coupled N=4 supersymmetric Yang-Mills
%plasma,\textquotedblright{}
Phys. Rev. Lett. 87, 081601 (2001) {[}arXiv:hep-th/0104066{]};
%\textquotedblleft{}From AdS/CFT correspondence to
%hydrodynamics,\textquotedblright{}
JHEP 0209, 043 (2002) {[}arXiv:hep-th/0205052{]}.

\bibitem{KSS}P. Kovtun, D.T. Son and A.O. Starinets, %\textquotedblleft{}Holography
%and hydrodynamics: Diffusion on stretched
%horizons,\textquotedblright{}
JHEP 0310, 064 (2003) {[}arXiv:hep-th/0309213{]}.

\bibitem{B-L}A. Buchel and J.T. Liu, %\textquotedblleft{}Universality
%of the shear viscosity in supergravity,\textquotedblright{}
Phys. Rev. Lett. 93, 090602 (2004) {[}arXiv:hep-th/0311175{]}.

\bibitem{I-L}N. Iqbal and H. Liu, %\textquotedblleft{}Universality
%of the hydrodynamic limit in AdS/CFT and the membrane
%paradigm,\textquotedblright{}
Phys. Rev. D 79, 025023 (2009) {[}arXiv:0809.3808{]}.

%\cite{Bhattacharyya:2008kq}
\bibitem{Bhattacharyya:2008kq}
   S.~Bhattacharyya, S.~Minwalla and S.~R.~Wadia,
  %``The Incompressible Non-Relativistic Navier-Stokes Equation from Gravity,''
  JHEP 0908, 059 (2009)
  [arXiv:0810.1545].
  %%CITATION = JHEPA,0908,059;%%

\bibitem{EFO} C. Eling, I. Fouxon and Y. Oz, %\textquotedblleft{}The
%Incompressible Navier-Stokes Equations From Membrane
%Dynamics,\textquotedblright{}
 Phys. Lett. B 680, 496 (2009) {[}arXiv:0905.3638{]}.

\bibitem{Padmanabhan10rp}
  T.~Padmanabhan,
  %Entropy density of spacetime and the Navier-Stokes fluid dynamics of null
  %surfaces,
  Phys. Rev. D 83, 044048 (2011) [arXiv:1012.0119].

\bibitem{Wilsonian}
I. Bredberg, C. Keeler, V. Lysov and A. Strominger, %Wilsonian
%Approach to Fluid/Gravity Duality,
JHEP 1103, 141 (2011) [arXiv:1006.1902].

%\cite{Heemskerk:2010hk}
\bibitem{Heemskerk10hk}
  I.~Heemskerk, J.~Polchinski,
  %``Holographic and Wilsonian Renormalization Groups,''
  JHEP 1106, 031 (2011) [arXiv:1010.1264].

%\cite{Faulkner:2010jy}
\bibitem{Faulkner10jy}
  T.~Faulkner, H.~Liu, M.~Rangamani,
  %``Integrating out geometry: Holographic Wilsonian RG and the membrane paradigm,''
  JHEP 1108, 051 (2011) [arXiv:1010.4036].

%\cite{Bredberg:2011jq}
\bibitem{Bredberg:2011jq}
  I.~Bredberg, C.~Keeler, V.~Lysov and A.~Strominger,
  %``From Navier-Stokes To Einstein,''
  JHEP {\bf 1207}, 146 (2012)
  [arXiv:1101.2451 [hep-th]].
  %%CITATION = ARXIV:1101.2451;%%

\bibitem{Compere:2011dx}
  G.~Compere, P.~McFadden, K.~Skenderis and M.~Taylor,
  %``The Holographic fluid dual to vacuum Einstein gravity,''
  JHEP {\bf 1107}, 050 (2011)
  [arXiv:1103.3022 [hep-th]].
  %%CITATION = ARXIV:1103.3022;%%

\bibitem{Cai}R.-G. Cai, L. Li and Y.-L. Zhang, %Non-Relativistic Fluid
%Dual to Asymptotically AdS Gravity at Finite Cutoff Surface,
JHEP 1107, 027 (2011) [arXiv:1104.3281].

%\cite{Bredberg:2011xw}
\bibitem{Bredberg:2011xw}
  I.~Bredberg and A.~Strominger,
  %``Black Holes as Incompressible Fluids on the Sphere,''
  JHEP {\bf 1205}, 043 (2012)
  [arXiv:1106.3084 [hep-th]].
  %%CITATION = ARXIV:1106.3084;%%

%\cite{Niu:2011gu}
\bibitem{Niu:2011gu}
  C.~Niu, Y.~Tian, X.~N.~Wu and Y.~Ling,
  %``Incompressible Navier-Stokes Equation from Einstein-Maxwell and Gauss-Bonnet-Maxwell Theories,''
  Phys.\ Lett.\ B {\bf 711}, 411 (2012)
  [arXiv:1107.1430 [hep-th]].
  %%CITATION = ARXIV:1107.1430;%%

%\cite{Matsuo:2011fk}
\bibitem{Matsuo11fk}
  S.~-J.~Sin, Y.~Zhou,
  %``Holographic Wilsonian RG Flow and Sliding Membrane Paradigm,''
  JHEP 1105, 030 (2011) [arXiv:1102.4477];
  Y.~Matsuo, S.~J.~Sin and Y.~Zhou,
  %``Mixed RG Flows and Hydrodynamics at Finite Holographic Screen,''
  JHEP {\bf 1201}, 130 (2012)
  [arXiv:1109.2698 [hep-th]].
  %%CITATION = ARXIV:1109.2698;%%

\bibitem{Lysov11xx}
  V.~Lysov and A.~Strominger,
  ``From Petrov-Einstein to Navier-Stokes,''
  arXiv:1104.5502.

\bibitem{HL}
T. Huang, Y. Ling, W. Pan, Y. Tian and X. Wu, %From Petrov-Einstein
%to Navier-Stokes in Spatially Curved Spacetime,
JHEP 1110, 079 (2011) [arXiv:1107.1464].

%\cite{Huang:2011kj}
\bibitem{Huang:2011kj}
  T.~Z.~Huang, Y.~Ling, W.~J.~Pan, Y.~Tian and X.~N.~Wu,
  %``Fluid/gravity duality with Petrov-like boundary condition in a spacetime with a cosmological constant,''
  Phys.\ Rev.\ D {\bf 85}, 123531 (2012)
  [arXiv:1111.1576 [hep-th]].
  %%CITATION = ARXIV:1111.1576;%%

\bibitem{Zhang:2012uy}
  C.~Y.~Zhang, Y.~Ling, C.~Niu, Y.~Tian and X.~N.~Wu,
  %``Magnetohydrodynamics from gravity,''
  Phys.\ Rev.\ D {\bf 86}, 084043 (2012)
  [arXiv:1204.0959 [hep-th]].
  %%CITATION = ARXIV:1204.0959;%%

%\cite{Wu:2013kqa}
\bibitem{Wu:2013kqa}
  X.~Wu, Y.~Ling, Y.~Tian and C.~Zhang,
  %``Fluid/Gravity Correspondence For General Non-rotating Black Holes,''
  Class.\ Quant.\ Grav.\  {\bf 30}, 145012 (2013)
  [arXiv:1303.3736 [hep-th]].
  %%CITATION = ARXIV:1303.3736;%%

%\cite{Wu:2013mda}
\bibitem{Wu:2013mda}
  B.~Wu and L.~Zhao,
  %``Gravity-mediated holography in fluid dynamics,''
  Nucl.\ Phys.\ B {\bf 874}, 177 (2013)
  [arXiv:1303.4475 [hep-th]].
  %%CITATION = ARXIV:1303.4475;%%

%\cite{Ling:2013kua}
\bibitem{Ling:2013kua}
  Y.~Ling, C.~Niu, Y.~Tian, X.~N.~Wu and W.~Zhang,
  %``Note on the Petrov-like boundary condition at finite cutoff surface in gravity/fluid duality,''
  Phys.\ Rev.\ D {\bf 90}, no. 4, 043525 (2014)
  [arXiv:1306.5633 [gr-qc]].
  %%CITATION = ARXIV:1306.5633;%%

%\cite{Cai:2013uye}
\bibitem{Cai:2013uye}
  R.~G.~Cai, L.~Li, Q.~Yang and Y.~L.~Zhang,
  %``Petrov type I Condition and Dual Fluid Dynamics,''
  JHEP {\bf 1304}, 118 (2013)
  [arXiv:1302.2016 [hep-th]].
  %%CITATION = ARXIV:1302.2016;%%

%\cite{Cai:2014ywa}
\bibitem{Cai:2014ywa}
  R.~G.~Cai, Q.~Yang and Y.~L.~Zhang,
  %``Petrov type I Spacetime and Dual Relativistic Fluids,''
  Phys.\ Rev.\ D {\bf 90}, no. 4, 041901 (2014)
  [arXiv:1401.7792 [hep-th]].
  %%CITATION = ARXIV:1401.7792;%%

%\cite{Cai:2014sua}
\bibitem{Cai:2014sua}
  R.~G.~Cai, Q.~Yang and Y.~L.~Zhang,
  %``Petrov type I Condition and Rindler Fluid in Vacuum Einstein-Gauss-Bonnet Gravity,''
  JHEP {\bf 1412}, 147 (2014)
  [arXiv:1408.6488 [hep-th]].
  %%CITATION = ARXIV:1408.6488;%%

%\cite{Hao:2014xva}
\bibitem{Hao:2014xva}
  X.~Hao, B.~Wu and L.~Zhao,
  %``Flat space compressible fluid as holographic dual of black hole with curved horizon,''
  JHEP {\bf 1502}, 030 (2015)
  [arXiv:1412.8144 [hep-th]].
  %%CITATION = ARXIV:1412.8144;%%

%\cite{Hao:2015zxa}
\bibitem{Hao:2015zxa}
  X.~Hao, B.~Wu and L.~Zhao,
  %``Compressible forced viscous fluid from product Einstein manifolds,''
  arXiv:1501.05146 [hep-th].
  %%CITATION = ARXIV:1501.05146;%%

%\cite{Kovtun:2004de}
\bibitem{Kovtun:2004de}
  P.~Kovtun, D.~T.~Son and A.~O.~Starinets,
  %``Viscosity in strongly interacting quantum field theories from black hole physics,''
  Phys.\ Rev.\ Lett.\  {\bf 94}, 111601 (2005)
  [hep-th/0405231].
  %%CITATION = HEP-TH/0405231;%%

%\cite{Rebhan:2011vd}
\bibitem{Rebhan:2011vd}
  A.~Rebhan and D.~Steineder,
  %``Violation of the Holographic Viscosity Bound in a Strongly Coupled Anisotropic Plasma,''
  Phys.\ Rev.\ Lett.\  {\bf 108}, 021601 (2012)
  [arXiv:1110.6825 [hep-th]].
  %%CITATION = ARXIV:1110.6825;%%

%\cite{Cheng:2014sxa}
\bibitem{Cheng:2014sxa}
  L.~Cheng, X.~H.~Ge and S.~J.~Sin,
  %``Anisotropic plasma with a chemical potential and scheme-independent instabilities,''
  Phys.\ Lett.\ B {\bf 734}, 116 (2014)
  [arXiv:1404.1994 [hep-th]].
  %%CITATION = ARXIV:1404.1994;%%

%\cite{Cheng:2014qia}
\bibitem{Cheng:2014qia}
  L.~Cheng, X.~H.~Ge and S.~J.~Sin,
  %``Anisotropic plasma at finite $U(1)$ chemical potential,''
  JHEP {\bf 1407}, 083 (2014)
  [arXiv:1404.5027 [hep-th]].
  %%CITATION = ARXIV:1404.5027;%%

%\cite{Ge:2014aza}
\bibitem{Ge:2014aza}
  X.~H.~Ge, Y.~Ling, C.~Niu and S.~J.~Sin,
  %``Holographic transports and stability in anisotropic linear axion model,''
  arXiv:1412.8346 [hep-th].
  %%CITATION = ARXIV:1412.8346;%%

%\cite{Jain:2014vka}
\bibitem{Jain:2014vka}
  S.~Jain, N.~Kundu, K.~Sen, A.~Sinha and S.~P.~Trivedi,
  %``A Strongly Coupled Anisotropic Fluid From Dilaton Driven Holography,''
  JHEP {\bf 1501}, 005 (2015)
  [arXiv:1406.4874 [hep-th]].
  %%CITATION = ARXIV:1406.4874;%%

%\cite{Critelli:2014kra}
\bibitem{Critelli:2014kra}
 {R.~Critelli, S.~I.~Finazzo, M.~Zaniboni and J.~Noronha,
  %``Anisotropic shear viscosity of a strongly coupled non-Abelian plasma from magnetic branes,''
  Phys.\ Rev.\ D {\bf 90}, 066006 (2014)
  [arXiv:1406.6019 [hep-th]].}
  %%CITATION = ARXIV:1406.6019;%%

  %\cite{Jain:2015txa}
\bibitem{Jain:2015txa}
  S.~Jain, R.~Samanta and S.~P.~Trivedi,
  %``The Shear Viscosity in Anisotropic Phases,''
  JHEP {\bf 1510}, 028 (2015)
  [arXiv:1506.01899 [hep-th]].
  %%CITATION = ARXIV:1506.01899;%%

%\cite{Friedrich:1998xt}
\bibitem{Friedrich:1998xt}
  H.~Friedrich and G.~Nagy,
  %``The Initial boundary value problem for Einstein's vacuum field equations,''
  Commun.\ Math.\ Phys.\  {\bf 201}, 619 (1999).

%\cite{Mateos:2011ix}
\bibitem{Mateos:2011ix}
  D.~Mateos and D.~Trancanelli,
  %``The anisotropic N=4 super Yang-Mills plasma and its instabilities,''
  Phys.\ Rev.\ Lett.\  {\bf 107}, 101601 (2011)
  [arXiv:1105.3472 [hep-th]].
  %%CITATION = ARXIV:1105.3472;%%

%\cite{Mateos:2011tv}
\bibitem{Mateos:2011tv}
  D.~Mateos and D.~Trancanelli,
  %``Thermodynamics and Instabilities of a Strongly Coupled Anisotropic Plasma,''
  JHEP {\bf 1107}, 054 (2011)
  [arXiv:1106.1637 [hep-th]].
  %%CITATION = ARXIV:1106.1637;%%

\bibitem{DG1}A. Donos and J.P. Gauntlett, ``Navier-Stokes on Black Hole Horizons and DC Thermoelectric Conductivity,'' [arXiv:1506.01360].

\bibitem{DG2}A. Donos and J.P. Gauntlett, ``Thermoelectric DC conductivities and Stokes flows on black hole horizons,'' [arXiv:1507.00234].


\end{thebibliography}
\end{document}